\documentclass[12pt]{article}

\usepackage{graphicx,cite}
\usepackage{epsf,latexsym}
\usepackage{amssymb,amsmath}
\usepackage{dcolumn}
\usepackage{bm}
\usepackage{txfonts}
\renewcommand{\baselinestretch}{1.5}

\begin{document}

\def\llra{\relbar\joinrel\longrightarrow}              
\def\mapright#1{\smash{\mathop{\llra}\limits_{#1}}}    
\def\mapup#1{\smash{\mathop{\llra}\limits^{#1}}}     
\def\mapupdown#1#2{\smash{\mathop{\llra}\limits^{#1}_{#2}}} 

\title{\bf Dense nuclear matter and symmetry energy in strong magnetic fields }

\author{Jianmin Dong$^{1,2,3}$, Umberto Lombardo$^{4}$, Wei Zuo$^{1}$,
   Hongfei Zhang $^{2}$ }

\maketitle
\begin{center}
\begin{enumerate}
\item Institute of Modern
Physics, Chinese Academy of Sciences, Lanzhou 730000, China
\item School of Nuclear Science and Technology, Lanzhou University, Lanzhou 730000, China
\item Graduate University of Chinese Academy of Sciences, Beijing 100049, China
\item Dipartimento di Fisica and INFN-LNS, Via S. Sofia 64, I-95123 Catania, Italy
\end{enumerate}
\end{center}

\maketitle

\begin{abstract}
\noindent The properties of nuclear matter in the presence of a
strong magnetic field, including the density-dependent symmetry
energy, the chemical composition and spin polarizations, are
investigated in the framework of the relativistic mean field models
FSU-Gold. The anomalous magnetic moments (AMM) of the particles and
the nonlinear isoscalar-isovector coupling are included. It is found
that the parabolic isospin-dependence of the energy per nucleon of
asymmetric nuclear matter remains valid for values of the magnetic
field below $10^{5}B_{c}^{e}$, $B_{c}^{e}=4.414\times10^{13}$G being
the electron critical field. Accordingly, the symmetry energy can be
obtained by the difference of the energy per nucleon in pure neutron
matter and that in symmetric matter. The symmetry energy, which is
enhanced by the presence of the magnetic field, significantly
affects the chemical composition and the proton polarization. The
effects of the AMM of each component on the energy per nucleon,
symmetry energy, chemical composition and spin polarization are
discussed in detail.

\end{abstract}



\maketitle
\section{Introduction}\label{intro}\noindent
Pulsars have been identified as rapidly rotating magnetized neutron
stars, with a surface magnetic field as large as $10^{11}-10^{13}$ G
\cite{HPY}. It is currently assumed that soft gamma repeaters and
anomalous X-ray pulsars have a strong surface magnetic fields, up to
$10^{14}-10^{15}$ G \cite{MAS}. The magnetic field in the interior
could be as large as $10^{18}$ G according to the scalar virial
theorem \cite{DL}. The energy of a charged particle changes
significantly if the magnetic field is comparable to or above a
critical value. This critical field is defined as that value where
the cyclotron energy is equal to the rest energy of the charged
particle, which for electrons is $B_{c}^{e}=4.414\times10^{13}$G.
This value is usually taken as the unit of the strong magnetic
field. In addition, the strong magnetic fields are also created in
heavy-ion collisions. For example, in noncentral Au + Au collisions
at 100 GeV/nucleon, the maximal magnetic field can reach about
$10^{17}$ G \cite{C1,C2}. The strong magnetic field is able to
affect the dynamics of heavy ion reactions \cite{C3}. Therefore, one
may expect considerable influence of such intense magnetic fields on
the properties of neutron star matter and of neutron star, in
particular the high density behavior of the symmetry energy.

Many investigations have been carried out on dense matter in strong
magnetic fields, such as the equation of states
\cite{SD,EoS1,EoS2,EoS3,EoS4,MAO,SH,NRT}, transport properties and
the cooling or heating of magnetized stars
\cite{Cool1,Cool2,YK,Cool3,Cool4,Cool5}. The gross properties of
cold symmetric matter and neutron star matter under the influence of
strong magnetic fields were investigated in the relativistic Hartree
theory in Ref. \cite{SD}. It was found that when the magnetic field
is about $10^{18}$ G and more, the nuclear matter in $\beta$
equilibrium practically converts into the stable proton-rich matter.
However, in these studies the effects of the anomalous magnetic
moments (AMM) of nucleons were completely neglected. In Ref.
\cite{MAO}, the authors based on relativistic mean field (RMF)
models argued that an extremely strong magnetic field may lead to
the pure neutron matter instead of the proton-rich matter when the
nucleon AMM are included. This indicates the importance of the AMM
effect. In addition, the structure of neutron stars can be
significantly affected by the strong magnetic fields
\cite{NS00,NS01,NS02,NS03}. The magnetic field actually influences
the neutron star structure in two ways: by modifying the energy
density and pressure of the neutron star matter, and itself provides
an additional energy density and pressure.

The energy per particle in nuclear matter is $e(\rho ,\beta )=e(\rho
,0)+S(\rho )\beta ^{2}+\mathcal{O(}\beta ^{4})$ with the density
$\rho =\rho _{n}+\rho _{p}$ and asymmetry $\beta =(\rho _{n}-\rho
_{p})/\rho $. The density-dependent symmetry energy $S(\rho )$ that
characterizes the isospin-dependent part of the equation of state
(EOS) of asymmetric nuclear matter plays a crucial role in
understanding a variety of issues in nuclear physics as well as
astrophysics, such as the heavy ion reactions
\cite{PD,AWS,VB,BAL,JML}, the stability of superheavy nuclei
\cite{JD}, and the structures, composition and cooling of neutron
stars \cite{NS1,NS,BG,DONG0}. Many theoretical and experimental
efforts have been made to constrain the density-dependent symmetry
energy. In this work, some properties of nuclear matter especially
the symmetry energy and its effects in magnetic fields are
investigated with a new relativistic mean field model proposed in
Ref. \cite{BG}. And not only the AMM of nucleons but also the one of
leptons are included, and the AMM of each component will be analyzed
in detail. The electron AMM perhaps is important because it is
larger than the proton and neutron magnetic moment owing to the
large Bohr magneton with respect to nuclear magneton. The presence
of a strong magnetic field could cause the polarization of nuclear
matter, an issue recently investigated in many theoretical works
\cite{SH,GM4,GM5}.

This paper is organized as follows: In Section 2, a brief framework
of the relativistic mean field model of the nuclear matter in a
uniform magnetic field $B$ is presented. Then the density dependent
symmetry energy, fractions and spin polarization affected by the
magnetic fields and AMM are shown in Sections 3, 4 and 5,
respectively. Finally, a brief summary is provided in Section 6.

\section{Brief framework of the method}\label{model}\noindent
In the RMF theory of nuclear matter that made of nucleons (p,n) and
leptons (e, $\mu$) in a uniform magnetic field $B$ along the $z$
axis, the total interacting Lagrangian density is given by
\begin{eqnarray}
\mathcal{L}&=&\overline{\psi }_{b}(i\gamma ^{\mu }\partial _{\mu
}-M-g_{\sigma
}\sigma -\frac{1}{2}g_{\rho }\gamma ^{\mu }{\bm\tau }\cdot {\bm\rho _{\mu }}  \notag \\
&&{-q}_{{b}}{\gamma ^{\mu }\frac{1+\tau _{3}}{2}A_{\mu
}-}\frac{1}{4}\kappa _{b}\sigma _{\mu \nu }F^{\mu \nu })\psi
_{b}-\frac{1}{4}F_{\mu \nu
}F^{\mu \nu } \notag\\
&&+\frac{1}{2}\partial _{\mu }\sigma \partial ^{\mu }\sigma -(\frac{1}{2}%
m_{\sigma }^{2}\sigma ^{2}+\frac{1}{3}g_{2}\sigma ^{3}+\frac{1}{4}%
g_{3}\sigma ^{4})  \notag \\
&&-\frac{1}{4}\Omega _{\mu \nu }\Omega ^{\mu \nu
}+\frac{1}{2}m_{\omega }^{2}\omega _{\mu }\omega ^{\mu }+\frac{\zeta
}{4!}g_{\omega }^{2}(\omega
_{\mu }\omega ^{\mu })^{2}  \notag \\
&&+g_{\omega }\overline{\psi }\gamma ^{\mu }\psi \omega _{\mu
}+\Lambda _{v}g_{\rho }^{2}{\rho }_{\mu }\cdot {\rho }^{\mu
}g_{\omega }^{2}\omega
_{\mu }\omega ^{\mu } \notag\\
&&-\frac{1}{4}{\bm R}_{\mu \nu }\cdot {\bm R}^{\mu \nu
}+\frac{1}{2}m_{\rho
}^{2}{\bm\rho }_{\mu }\cdot {\bm\rho }^{\mu }  \notag \\
&&+\overline{\psi }_{l}(i\gamma ^{\mu }\partial _{\mu }-m_{l}-{q}_{{l}}{%
\gamma ^{\mu }A_{\mu }-}\frac{1}{4}\kappa _{l}\sigma _{\mu \nu
}F^{\mu \nu })\psi _{l}
\end{eqnarray}
with $A^{\mu }=(0,0,Bx,0)$ and $\sigma ^{\mu \nu }=\frac{i}{2}\left[
\gamma ^{\mu },\gamma ^{\nu }\right] $. $\mu _{N}$ ($\mu _{B}$)
denotes the nuclear (Bohr) magneton of nucleons (leptons). $\kappa
_{p}=1.7928\mu _{N}$, $\kappa _{n}=-1.9130\mu _{N}$, $\kappa
_{e}=1.15965\times 10^{-3}\mu _{B}$ and $\kappa _{\mu}=1.16592\times
10^{-3}\mu _{B}$ are the AMM for protons, neutrons, electrons and
muons, respectively~\cite{mass}. $M$, $m_{\sigma }$, $m_{\omega }$
and $m_{\rho }$ are the nucleon-, the $\sigma $-, the $\omega$- and
the $\rho $-meson masses, respectively. The nucleon field $\psi_{b}
$ interacts with the $\sigma ,\omega ,\rho $ meson fields $\sigma
,\omega _{\mu },\rho _{\mu }$ and with the photon field $A_{\mu }$.
The field tensors for the vector meson are given as $\Omega _{\mu\nu
}=\partial _{\mu }\omega _{\nu }-\partial _{\nu }\omega _{\mu }$ and
by similar expression for $\rho $ meson and the photon. The
self-coupling terms with coupling constants $g_{2}$ and $g_{3}$ for
the $\sigma $ meson are introduced because they turned out to be
crucial~\cite{BB}. Compared with the previous RMF models, the RMF
interactions employed in this work is FSUGold where two additional
parameters $\zeta $ and $\Lambda _{v}$ have been introduced:
$\omega$ meson self-interactions as described by $\zeta $ which
soften the equation of state at high density, and the nonlinear
mixed isoscalar-isovector coupling described by $\Lambda _{v}$ that
modifies the density-dependence of the symmetry energy. The FSUGold
interaction \cite{BG} gives a good description of ground state
properties as well as excitations of finite nuclei.

The energy densities of proton, neutron, electron and muon are given
by
\begin{equation}
\varepsilon _{p}=\frac{eB}{4\pi ^{2}}\underset{\nu ,s}{\sum }\left[
k_{f,\nu ,s}^{p}E_{f}^{p}+\left( \sqrt{M^{\ast 2}+2\nu eB}-s\kappa
_{p}B\right)
^{2}\ln \left\vert \frac{k_{f,\nu ,s}^{p}+E_{f}^{p}}{\sqrt{M^{\ast 2}+2\nu eB%
}-s\kappa _{p}B}\right\vert \right],
\end{equation}%

\begin{eqnarray}
\varepsilon _{n} &=&\frac{1}{4\pi ^{2}}\underset{s}{\sum }\bigg\{\frac{1}{2}%
k_{f,s}^{n}E_{f}^{n3}-\frac{2}{3}s\kappa _{n}BE_{f}^{n3}\left( \arcsin \frac{%
M^{\ast }-s\kappa _{n}B}{E_{f}^{n}}-\frac{\pi }{2}\right) -\left( \frac{%
s\kappa _{n}B}{3}+\frac{M^{\ast }-s\kappa _{n}B}{4}\right) \times  \nonumber\\
&&\left[ (M^{\ast }-s\kappa _{n}B)k_{f,s}^{n}E_{f}^{n}+(M^{\ast
}-s\kappa
_{n}B)^{3}\ln \left\vert \frac{k_{f,n}^{p}+E_{f}^{p}}{M^{\ast }-s\kappa _{p}B%
}\right\vert \right] \bigg\},
\end{eqnarray}

\begin{equation}
\varepsilon _{e}=\frac{eB}{4\pi ^{2}}\underset{\nu ,s}{\sum }\left[
k_{f,\nu ,s}^{e}E_{f}^{e}+\left( \sqrt{m_{e}^{2}+2\nu eB}-s\kappa
_{e}B\right)
^{2}\ln \left\vert \frac{k_{f,\nu ,s}^{e}+E_{f}^{e}}{\sqrt{m_{e}^{2}+2\nu eB}%
-s\kappa _{e}B}\right\vert \right],
\end{equation}%
\begin{equation}
\varepsilon _{\mu }=\frac{eB}{4\pi ^{2}}\underset{\nu ,s}{\sum
}\left[ k_{f,\nu ,s}^{\mu }E_{f}^{\mu }+\left( \sqrt{m_{\mu
}^{2}+2\nu eB}-s\kappa
_{\mu }B\right) ^{2}\ln \left\vert \frac{k_{f,\nu ,s}^{\mu }+E_{f}^{\mu }}{%
\sqrt{m_{\mu }^{2}+2\nu eB}-s\kappa _{\mu }B}\right\vert \right],
\end{equation}
where $\nu=0,1,2,3...$ denotes the Landau levels for charged
particles. The summation of $\nu$ starts from 0(1) for spin-up
protons (leptons), and $\nu$ runs up to largest integer for which
the Fermi momentum squared $k_{f,\nu ,s}^{2}$ is positive. The
energy density of neutron star matter can be expressed as
\begin{equation}
\varepsilon =\varepsilon _{p}+\varepsilon _{n}+\varepsilon
_{e}+\varepsilon
_{\mu }+\frac{1}{2}m_{\sigma }^{2}\sigma ^{2}+\frac{1}{3}g_{2}\sigma ^{3}+%
\frac{1}{4}g_{3}\sigma ^{4}+\frac{1}{2}m_{\omega }^{2}\omega _{0}^{2}+\frac{1%
}{2}m_{\rho }^{2}\rho _{30}^{2}+3\Lambda _{v}g_{\rho }^{2}\rho
_{30}^{2}g_{\omega }^{2}\omega _{0}^{2}+\frac{\zeta }{8}g_{\omega
}^{2}\omega _{0}^{4}.
\end{equation}
The nucleons and leptons satisfy the chemical equilibrium condition
$\mu_{n}-\mu_{p}=\mu_{e}=\mu_{\mu}$ and charge neutrality condition
$\rho_{p}=\rho_{e}+\rho_{\mu}$.

\section{Density dependent symmetry energy}\label{model}\noindent
First of all, we investigate the properties of nuclear matter
(without leptons) in magnetic field with the inclusion of the
nucleon AMM. As shown in Fig. 1, the energy per nucleon of
asymmetric nuclear matter versus $\beta^{2}$ still fulfills the
parabolic law for the selected densities, i.e., $\rho_{0}$,
$2\rho_{0}$, $3\rho_{0}$ and $5\rho_{0}$, where $\rho_{0}=0.16$
fm$^{-3}$ is the saturation density. As expected, increasing the
magnetic field the energy per particle turns out to be increasing
lower, especially at low densities. In fact, the Landau quantization
of the charged particles causes a softening of the equation of state
(EOS), as already found in Ref. \cite{SH}. At large densities such
as $\rho=5\rho_{0}$, as the softening of the EOS by the Landau
quantization will be gradually overwhelmed by the stiffening
resulting from the AMM effect \cite{LT,ACJ}, the energies are
slightly reduced by a strong magnetic field. In addition, it is
found that the more neutron-rich the dense matter is, the weaker of
changes the magnetic field causes. This is simply explained by the
fact that neutrons carry no charge so that they have no Landau
levels to fill. The direct coupling of neutrons to magnetic field is
just due to the neutron AMM. For protons, however, their charge
strongly couples with the magnetic field, forming the Landau levels,
and this coupling is much stronger than the direct coupling between
the AMM and magnetic field. Since the $\beta^{2}$ law still holds in
magnetic field below $10^{5}B_{c}^{e}$, the symmetry energy can be
obtained just as the field-free case. The influences of the proton
and neutron AMM on the energy per nucleon $E/A$ is illustrated in
Fig. 2. The nucleon AMM lead to a reduction of $E/A$ due to their
coupling to the magnetic field. At low isospin asymmetry $\beta$,
the proton AMM contributes more sizeably than the neutron AMM,
because the coupling between the proton AMM and the magnetic field
leads to a non-negligible change in the filling of Landau levels and
hence the energy spectra. At high isospin asymmetry, the effect of
the neutron AMM is larger than that of the proton AMM because of the
large neutron fraction.

In the upper panel of Fig. 3, we plot the density dependence of the
symmetry energy for various magnetic field strengths with the
inclusion of AMM. The employed RMF interaction is FSU-Gold
\cite{BG}. Other interactions with some modifications of $\Lambda
_{v}$ and $g_{\rho}$ based on FSU-Gold \cite{BS} do not lead to very
different results. The density-dependent symmetry energy is enhanced
in strong magnetic field, in particular at low densities. Whereas
the effect of the magnetic field is rather weak below
$5\times10^{4}B_{c}^{e}$. As we mentioned above, the energy per
particle of the neutron is reduced more slightly than that of
proton. Therefore, the symmetry energy, defined as the energy per
particle difference between pure neutron matter and symmetric
matter, gets enhanced, with the effect of changing the fraction of
each component. To show the shift of the symmetry energy due to the
nucleon AMM, we report in the lower panel of Fig. 3 the symmetry
energy of the nuclear matter in a magnetic field with and without
the AMM in calculations. It can be seen that the effect of the
proton AMM is what we expected while the one of the neutron AMM is
much weaker, as one can easily realize from the results of Fig. 2
and the corresponding comments.

\section{Fractions of each component in the neutron star matter}\label{model}\noindent
The proton fractions $Y_{p}$ and muon fractions $Y_{\mu}$ for
$\beta$-stable dense matter are displayed in Fig. 4 as a function of
the magnetic field strength under different densities. The neutron
and electron fractions are obtained via the relations
$Y_{n}=1-Y_{p}$ and $Y_{e}=Y_{p}-Y_{\mu}$. The interactions with
some modifications of $\Lambda _{v}$ and $g_{\rho}$ based on
FSU-Gold \cite{BS} which scan different behavior of symmetry energy
from very stiff ($\Lambda _{v}=0.00$) to very soft ($\Lambda
_{v}=0.04$), are employed in calculations. It can be found that the
proton and muon fractions remain unaltered compared with the
field-free case for relatively weak magnetic fields. At the
saturation density, the proton fractions (and muon fractions) are
not very different from each other, because the symmetry energy
either soft or stiff, takes a value in agreement with the empirical
one. However, with increasing density, this difference becomes
sizable. The stiffer the symmetry energy is, the larger the proton
and muon fractions are. As soon as the magnetic field becomes strong
enough, such as $B>2\times10^{4}B_{c}^{e}$, those fractions depend
obviously not only on the symmetry energy but also on the magnetic
field strength at a given density. And the proton and muon fractions
rapidly bend up starting from a certain value of the magnetic field.
Beyond this threshold, the charged particles are completely spin
polarized due to Landau quantization with the Fermi energies
considerably reduced. This threshold is, for instance,
$B=5\times10^{4}B_{c}^{e}$ at $\rho=0.16$ fm$^{-3}$ for the proton
fraction. At the higher density matter, a much stronger magnetic
field is required to reach the threshold. Besides, the interaction
giving a stiff symmetry energy displays a larger threshold, which
stems from the fact that the stiff symmetry energy provides a large
proton fraction and hence a high proton density at a given nucleon
density. In a word, the fraction of each component of the neutron
star matter depends on both the density-dependent symmetry energy
and the magnetic field strength. For the field-free case, the
threshold for URCA process is that the proton fraction reaches about
$11\%$ at low densities. At high densities, when muons are involved,
this threshold becomes density dependent and can reach about $14\%$
\cite{URCA0}. But for the dense matter in magnetic field, as show in
Ref. \cite{YK}, the magnetic field smears out the threshold between
the open and closed direct URCA regimes producing magnetic
broadening of the direct URCA threshold. When $p_{Fn}< p_{Fp}+
p_{Fe}$, the direct URCA process is open, and when $p_{Fn} > p_{Fp}+
p_{Fe}$ that is forbidden at field free case the direct URCA process
can be still open if magnetic field is included. In order to study
the effects of the AMM of each particles on the proton and muon
fractions, Fig. 5 presents the $Y_{p}$ and $Y_{\mu}$ as a function
of magnetic field strength with and without the AMM, taking the
$\beta$-stable matter at $\rho=\rho_{0}$ as an example. The proton
AMM gives the biggest contribution compared to the AMM of other
constituents. The muon AMM almost has no effect on the fraction of
each component because of its large mass (small magnetic moment) and
low fraction. The electron and neutron AMM affect the fractions at
rather strong magnetic field with nearly the same amplitude, which
is not obvious. The effect of the proton AMM is evident because it
is able to affect the filling of Landau levels and hence the
corresponding fraction.

\section{Spin polarization of nucleons in neutron star matter}\label{model}\noindent
It is worthwhile discussing the spin polarization of neutrons and
protons in $\beta$-stable neutron star matter under the strong
magnetic field. The spin polarization of protons and neutrons in
fact can influence the superfluidity of neutron stars and, as a
consequence, its rotational dynamics and the cooling. The nucleon
spin-polarization is defined as
\begin{equation}
S_{\tau}=\frac{\rho _{\tau\uparrow }-\rho _{\tau\downarrow }}{\rho
_{\tau}},
\end{equation}%
with $\tau=p,n$. When no AMM is included, only proton spin
polarization may occur because the coupling between charge and
magnetic field survives and obviously affects the single particle
spectrum. At the critical field, all protons occupy the first Landau
level so as to reach full spin polarization. But when their AMM are
included, the energy spectra for both protons and neutrons are spin
dependent, and hence the neutrons can also experience spin
polarization for the coupling between their AMM and the magnetic
field. Since this coupling is much weaker than that between charged
particle and magnetic field, the complete polarization of neutrons
requires a much stronger magnetic field, as discussed in Ref.
\cite{SH}. Fig. 6 illustrates the spin polarization of protons and
neutrons with the inclusion of AMM adopting the same interactions as
in Fig. 4. The behavior of $S_p$ and $S_n$  results from the
interplay between the magnetic field intensity and the density
dependent symmetry energy. Both increase (in absolute value) as
increasing magnetic field for any density. At a larger density, it
requires a stronger magnetic field for proton to be fully spin
polarized. In other words, the proton becomes more difficult to be
polarized as the density increases. In addition, the larger the
symmetry energy is, the less protons can be polarized. This
phenomenon can be explained as follows. A larger symmetry energy may
lead to a less neutron-rich dense matter, namely, a larger fractions
of protons in $\beta$-stable matter. Yet, the protons in higher
density is more difficult to be polarized, as shown in upper panel
in Fig. 6. The spin polarization of neutrons is almost insensitive
to the symmetry energy, owing to the weak coupling between neutrons
and magnetic field.

Fig. 7 displays the AMM effect on the spin polarizations of protons
(upper panel) and neutron (lower panel) taking the $\beta$-stable
matter at $\rho=\rho_{0}$ as an example. The results with only the
lepton AMM and without any AMM completely coincide with each other,
indicating that the lepton AMM do not affect the spin polarizations
of nucleons. The proton AMM contributes most dominantly to the spin
polarizations of the proton since this AMM affects the filling of
Landau levels for protons. The neutron, if their AMM are neglected,
can not be polarized at all as shown in the lower panel. After the
AMM of neutron is taken into account, the doubly degeneracy with
opposite spin projections is destroyed and hence the neutrons show
the spin polarization.

\section{Summary}\label{sec5}\noindent

The properties of dense matter in strong magnetic fields, including
the density- dependent symmetry energy, chemical composition and
spin polarizations have been studied within the relativistic mean
field model FSU-Gold. The magnetic field does not modify the
parabolic behavior of the energy per particle in asymmetric matter
at least in the range of the magnetic field considered in this work,
and hence the symmetry energy can be derived just as the field-free
case. It is found that the strong magnetic field leads to an
enhancement of the symmetry energy with respect to field free case,
in particular at low densities. The fraction of each component has
been calculated with some modified FSU-Gold interactions that can
provide both stiff and soft symmetry energy. Once the magnetic field
is strong enough, the fractions of components are sensitive to both
the symmetry energy and the strong magnetic field. The
density-dependent symmetry energy affects the proton polarization
but it has almost no effect on neutron polarization. Moreover, the
effects of the AMM on the energy per nucleon, the symmetry energy,
fractions and spin-polarization were analyzed. The corresponding
effects on the symmetry energy and the fractions of each component
almost determined by the proton AMM, and the spin polarizations of
the protons (neutrons) are dominated by the proton (neutron) AMM.

\section*{Acknowledgements}
This work was supported by the Major State Basic Research Developing
Program of China under No. 2013CB834405, the National Natural
Science Foundation of China (with Grant Nos. 11175219,
10975190,11275271,11075066,11175074), the Knowledge Innovation
Project (KJCX2-EW-N01) of Chinese Academy of Sciences, Chinese
Academy of Sciences Visit- ing Professorship for Senior
International Scientists (Grant No.2009J2-26), CAS/SAFEA
International Partnership Program for Creative Research Teams
(CXTD-J2005-1), and the Funds for Creative Research Groups of China
under Grant No. 11021504.

\renewcommand{\baselinestretch}{1.0}

\newpage
\begin{figure}[htbp]
\begin{center}
\includegraphics[width=0.95\textwidth]{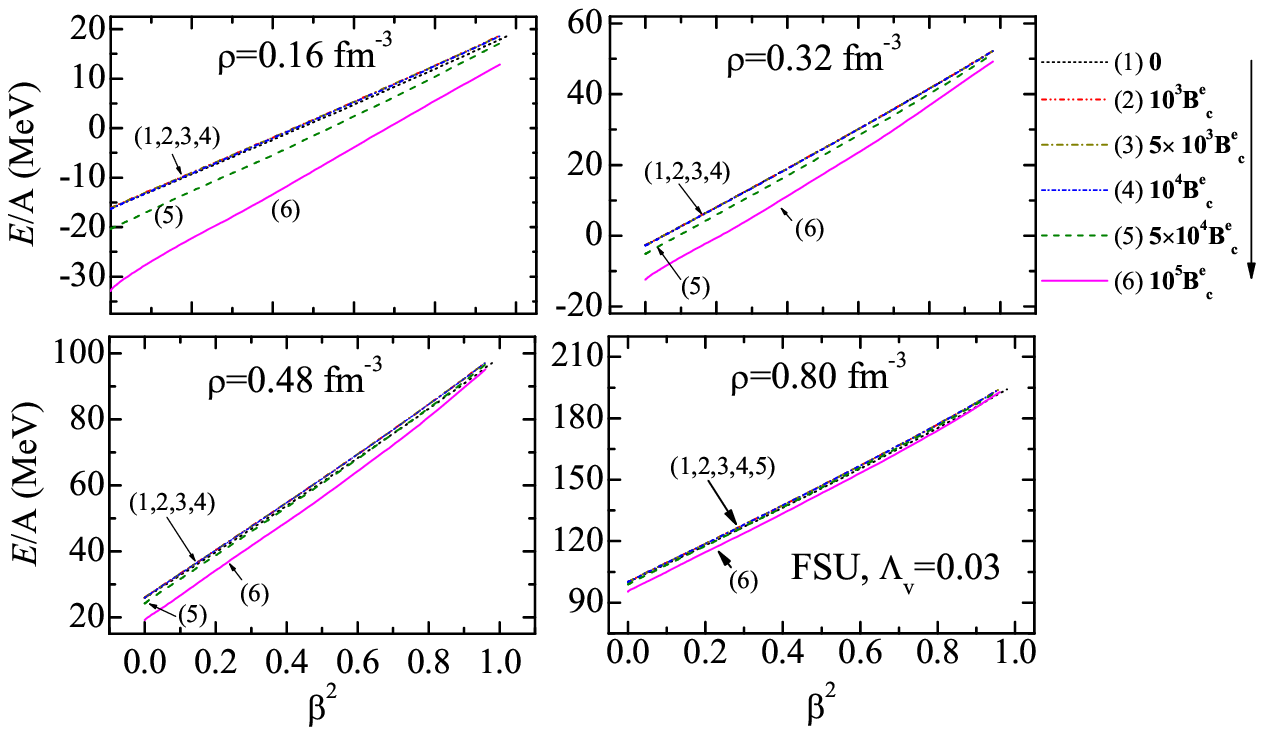}
\caption{(Color online) The energy per nucleon $E/A$ as a function
of $\beta^{2}$ for different values of the magnetic field $B$, where
$\beta$ is asymmetry. The RMF interaction FSU-Gold with $\Lambda
_{v}=0.03$ is used.}
\end{center}
\end{figure}

\newpage
\begin{figure}[htbp]
\begin{center}
\includegraphics[width=0.8\textwidth]{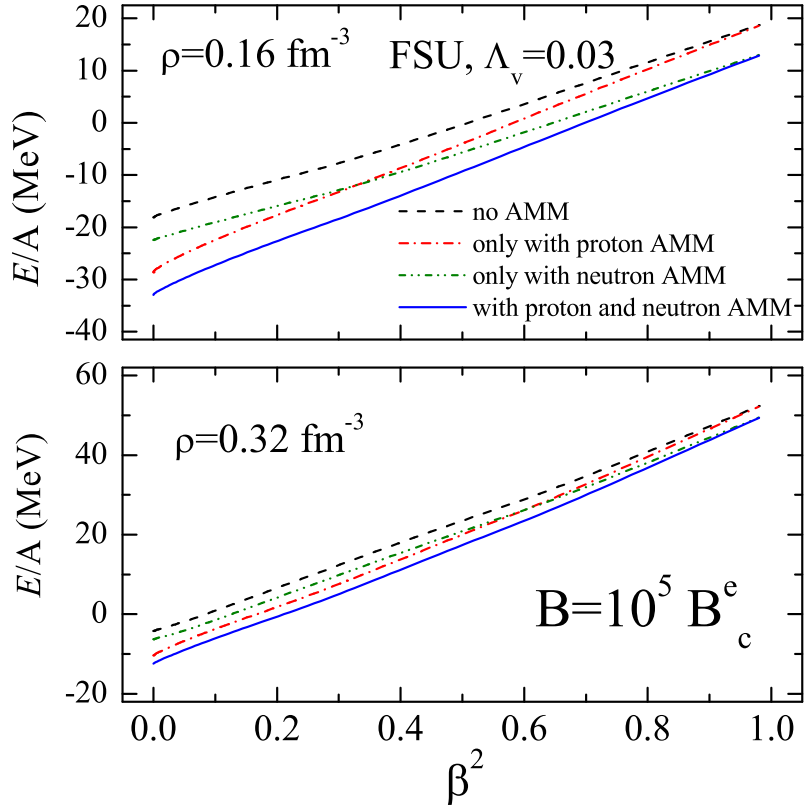}
\caption{(Color online) Effects of the nucleon AMM on the energy per
nucleon $E/A$ of asymmetric nuclear matter in a magnetic field of
$10^{5}B_{c}^{e}$. The interaction FSU-Gold with $\Lambda _{v}=0.03$
is used.}
\end{center}
\end{figure}

\newpage
\begin{figure}[htbp]
\begin{center}
\includegraphics[width=0.8\textwidth]{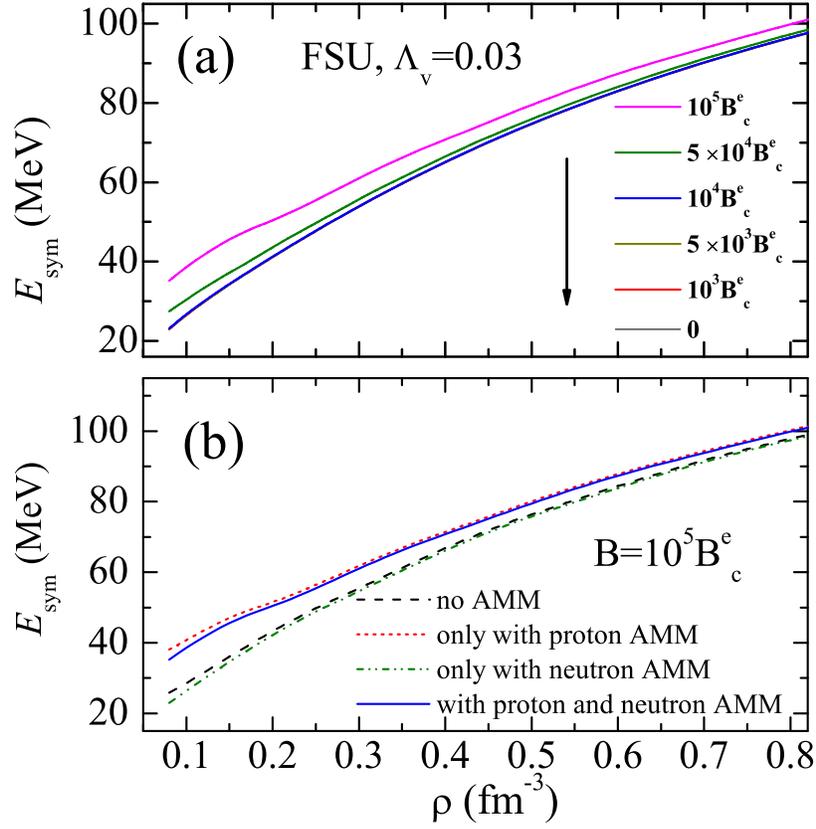}
\caption{(Color online) (a) Density dependence of the nuclear
symmetry energy for different magnetic field strengths. (b) Effects
of the AMM on the symmetry energy of nuclear matter in a magnetic
field of $10^{5}B_{c}^{e}$. The interaction FSU-Gold with $\Lambda
_{v}=0.03$ is used.}
\end{center}
\end{figure}

\newpage
\begin{figure*}[htbp]
\begin{center}
\includegraphics[width=0.8\textwidth]{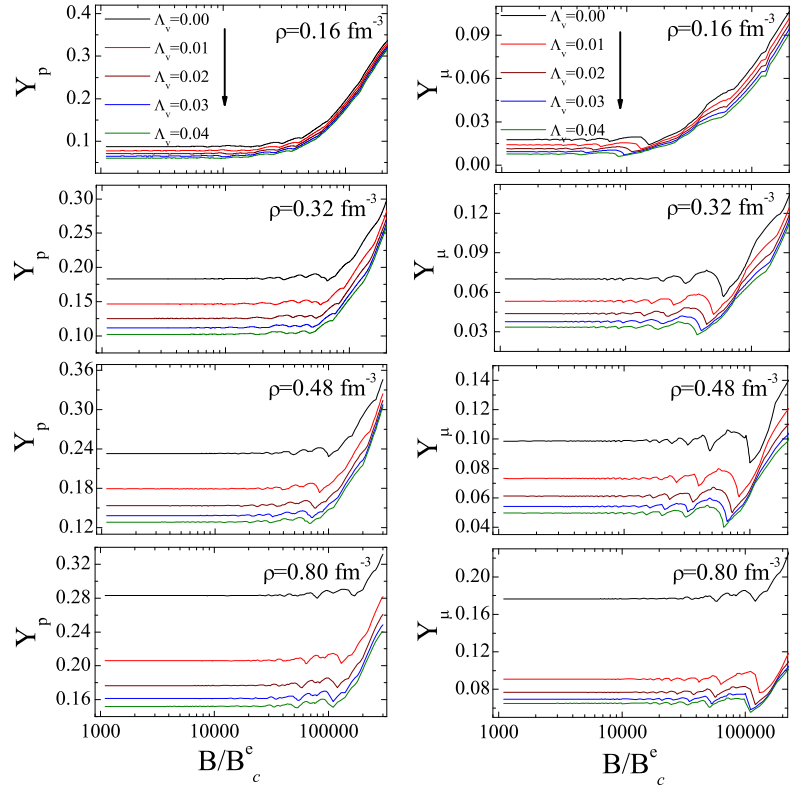}
\caption{(Color online) Proton and muon fractions in $\beta$-stable
matter versus the magnetic field strength $B$ under different
density $\rho$. The calculations are performed with modified
FSU-Gold interactions \cite{BS} providing stiff ($\Lambda
_{v}=0.00$) to soft ($\Lambda _{v}=0.04$) symmetry energy.}
\end{center}
\end{figure*}

\newpage
\begin{figure*}[htbp]
\begin{center}
\includegraphics[width=0.8\textwidth]{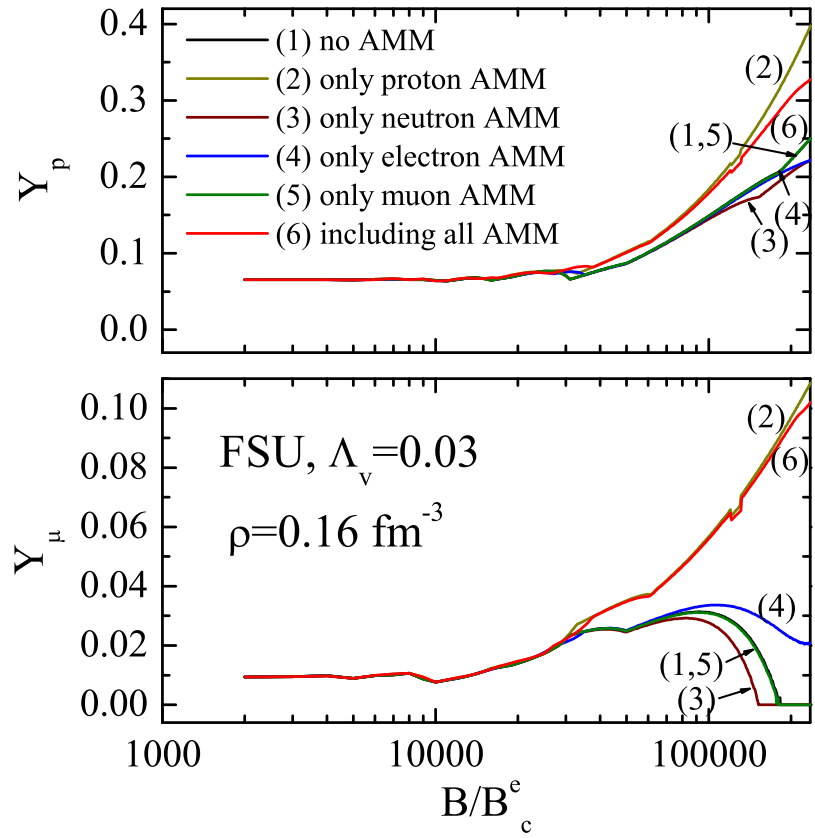}
\caption{(Color online) Effect of the AMM of each component on the
proton and muon fractions in $\beta$-stable matter versus the
magnetic field strength with the interactions FSU-Gold ($\Lambda
_{v}=0.03$), taking the density $\rho=\rho_{0}$ as an example.}
\end{center}
\end{figure*}

\newpage
\begin{figure*}[htbp]
\begin{center}
\includegraphics[width=0.8\textwidth]{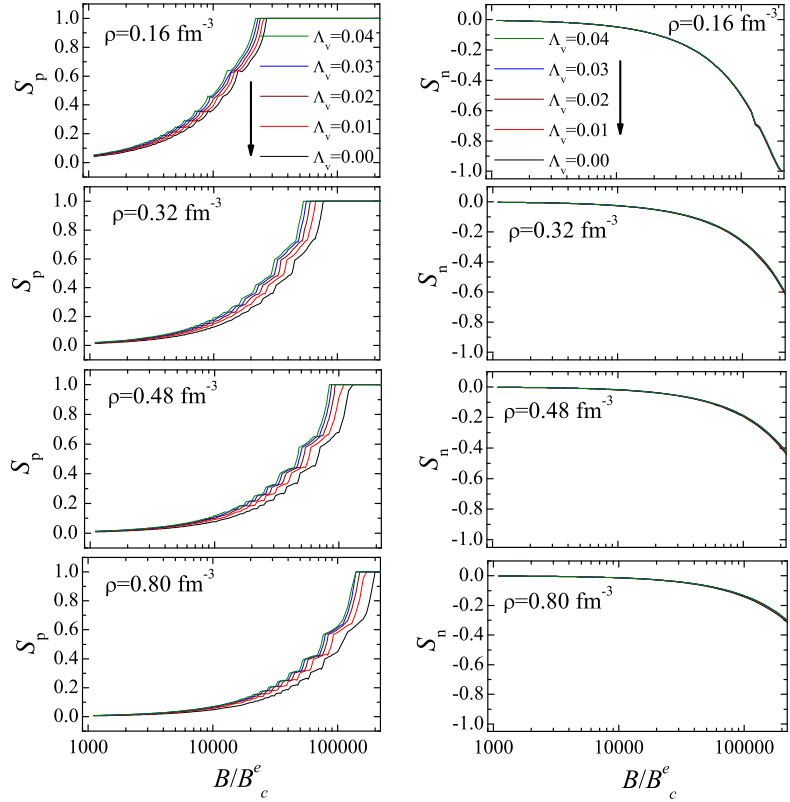}
\caption{(Color online) Spin polarization of protons and neutrons in
$\beta$-stable matter as a function of the magnetic field strength
$B$ under different density $\rho$. The calculations are performed
with modified FSU-Gold interactions \cite{BS} providing stiff
($\Lambda _{v}=0.00$) to soft ($\Lambda _{v}=0.04$) symmetry
energy.}
\end{center}
\end{figure*}

\newpage
\begin{figure*}[htbp]
\begin{center}
\includegraphics[width=0.8\textwidth]{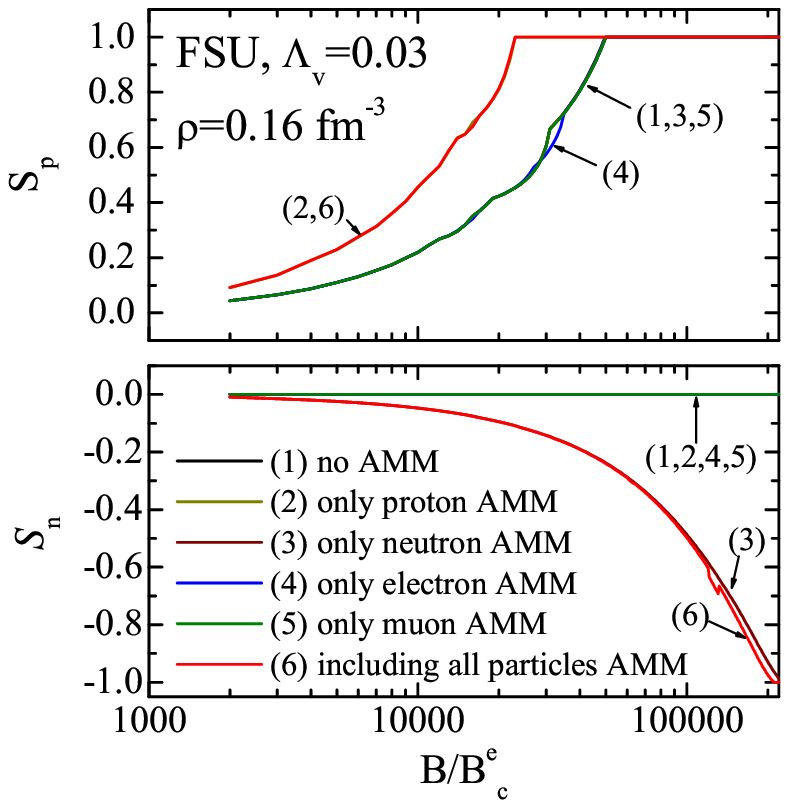}
\caption{(Color online) Effect of the AMM of each component on the
spin polarization of protons and neutrons as a function of the
magnetic field strength with the interactions FSU-Gold $\Lambda
_{v}=0.03$, taking the $\beta$-stable matter at the density
$\rho=\rho_{0}$ as an example.}
\end{center}
\end{figure*}
\bigskip

\begin{thebibliography}{00}

\bibitem{HPY}
P. Haensel, A. Y. Potekhin, and D. G. Yakovlev, Neutron Stars 1,
(2006).

\bibitem{MAS}
R. Duncan and C. Thompson, Astrophys. J. 392 (1992) L9; B.
Paczynski, Acta Astron. 42 (1992) 145; C. Kouveliotou et al., Nature
393 (1998) 235; A. Melatos, Astrophys. J. Lett. 519 (1999) L77.

\bibitem{DL}
D. Lai and S. L. Shapiro, Astrophys. J. 383 (1991) 745, and
references therein.


\bibitem{C1}
D. E. Kharzeev, L. D. McLerran, and H. J.Warringa, Nucl. Phys. A 803
(2008) 227.

\bibitem{C2}
V. Skokov, A. Illarionov, and V. Toneev, Int. J. Mod. Phys. A 24
(2009) 5925.

\bibitem{C3}
Li Ou, and Bao-An Li, Phys. Rev. C 84 (2011) 064605.


\bibitem{SD}
Somenath Chakrabarty, Debades Bandyopadhyay, and Subrata Pal, Phys.
Rev. Lett. 78 (1997) 2898.

\bibitem{EoS1}
A. E. Broderick, M. Prakash, J.M. Lattimer, Phys. Lett. B 531 (2002)
167.

\bibitem{EoS2}
F. X. Wei, G. J. Mao, C. M. Ko, L. S. Kisslinger, H. St\"ocker and
W. Greiner, J. Phys. G: Nucl. Part. Phys. 32 (2006) 47.

\bibitem{EoS3}
Efrain J. Ferrer, Vivian de la Incera, Jason P. Keith, Israel
Portillo, and Paul L. Springsteen, Phys. Rev. C 82(2010) 065802.

\bibitem{EoS4}
 A. A. Isayev and J. Yang, Phys. Rev. C 84 (2011) 065802; Phys. Lett. B707 (2012) 163.

\bibitem{MAO}
G. Mao, N. V. Kondratyev, A. Iwamoto, Z. Li, X. Wu, W. Greiner, and
N. I. Mikhailov, Chin. Phys. Lett. 20 (2003) 1238.

\bibitem{SH}
P. Yue and H. Shen, Phys. Rev. C 74 (2006) 045807.

\bibitem{NRT}
N. Chamel, R. L. Pavlov, L. M. Mihailov, Ch. J. Velchev, Zh. K.
Stoyanov, Y. D. Mutafchieva, M. D. Ivanovich, J. M. Pearson, S.
Goriely, arXiv:1210.5874v1 [astro-ph.HE] (2012).

\bibitem{Cool1}
V. G. Bezchastnov and P. Haensel, Phys. Rev. D 54 (1996) 3706.

\bibitem{Cool2}
Y. A. Shibanov and D. G. Yakovlev, Astron. Astrophys. 309 (1996)
171.

\bibitem{YK}
D. G. Yakovlev, A. D. Kaminker, O. Y. Gnedin, P. Haensel, Phys. Rep.
354 (2001) 1, and references therein.

\bibitem{Cool3}
T. Maruyama, T. Kajino, N. Yasutake, M. K. Cheoun, and C. Y. Ryu,
Phys. Rev. D 83 (2011) 081303(R).

\bibitem{Cool4}
Jos\'e A. Pons, Bennett Link, Juan A. Miralles, and Ulrich Geppert,
Phys. Rev. Lett. 98 (2007) 071101.

\bibitem{Cool5}
Deborah N. Aguilera, Vincenzo Cirigliano, Jos\'e A. Pons, Sanjay
Reddy, and Rishi Sharma, Phys. Rev. Lett. 102 (2009) 091101.


\bibitem{NS00}
Y. F. Yuan and J. L. Zhang, Astrophys. J. 525 (1999) 950.

\bibitem{NS01}
D. H. Wen, W. Chen and L. G. Liu, Commun. Theor. Phys. 47 (2007)
653.

\bibitem{NS02}
A. Rabhi, H. Pais, P. K. Panda and C. Providencia,J. Phys. G: Nucl.
Part. Phys. 36 (2009) 115204.

\bibitem{NS03}
A. Rabhi, P. K. Panda, and C. Providencia, Phys. Rev. C 84 (2011)
035803.


\bibitem{BG}
B. G. Todd-Rutel and J. Piekarewicz, Phys. Rev. Lett. 95 (2005)
122501.


\bibitem{PD}
P. Danielewicz, R. Lacey, and W. G. Lynch, Science 298,(2002) 1592.

\bibitem{AWS}
A. W. Steiner, M. Prakash, J. Lattimer, and P. J. Ellis, Phys. Rep.
411 (2005) 325.

\bibitem{VB}
V. Baran, M. Colonna, V. Greco, and M. Di Toro, Phys. Rep. 410
(2005) 335.

\bibitem{BAL}
B. A. Li, L. W. Chen, and C. M. Ko, Phys. Rep. 464 (2008) 113 .

\bibitem{JML}
J. M. Lattimer and M. Prakash, Phys. Rep. 442 (2007) 109.

\bibitem{JD}
Jianmin Dong, Wei Zuo, and Werner Scheid, Phys. Rev. Lett. 107
(2011) 012501.

\bibitem{NS1}
C. J. Horowitz and J. Piekarewicz, Phys. Rev. Lett. 86 (2001) 5647 .

\bibitem{NS}
J. M. Lattimer and M. Prakash, Phys. Rep. 333 (2000) 121; Science
304 (2004) 536.


\bibitem{DONG0}
Jianmin Dong, Wei Zuo, Jianzhong Gu, and Umberto Lombardo, Phys.
Rev. C 85 (2012) 034308.


\bibitem{GM4}
M. \'Angeles P\'erez-Garcia, Phys. Rev. C 77 (2008) 065806; Phys.
Rev. C 80 (2009) 045804.

\bibitem{GM5}
R. Aguirre, Phys. Rev. C 83 (2011) 055804.


\bibitem{mass}
K. Nakamura et al., (Particle Data Group), J. Phys. G: Nucl. Part.
Phys. 37 (2010) 075021.

\bibitem{BB}
J. Boguta and A. R. Bodmer, Nucl. Phys. A 292 (1977) 413.

\bibitem{LT}
A. Broderick, M. Prakash and J. M. Lattimer, Astrophys. J. 525
(1999) 950.

\bibitem{ACJ}
A. Rabhi, C. Providecia and J. Da. Providencia, J. Phys. G: Nucl.
Part. Phys. 35 (2008) 125201.

\bibitem{BS}
Bharat K. Sharma, Subrata Pal, Phys. Lett. B682 (2009) 23.

\bibitem{URCA0}
James M. Lattimer, C. J. Pethick, Madappa Prakash, Pawel Haensel,
Phys. Rev. Lett. 66 (1991) 2701.

\end{thebibliography}
\end{document}